\documentclass[conference,10pt]{IEEEtran}
\usepackage{cite}
\usepackage{amsmath,amssymb,amsfonts}
\usepackage{algorithmic}
\usepackage{graphicx}
\usepackage{textcomp}
\usepackage{xcolor}
\def\BibTeX{{\rm B\kern-.05em{\sc i\kern-.025em b}\kern-.08em
    T\kern-.1667em\lower.7ex\hbox{E}\kern-.125emX}}

\usepackage{bm}

\newcommand\Vector[1]{\bm{#1}}

\newcommand\0{{\Vector{0}}}

\newcommand\vh{{\Vector{h}}}

\newcommand\vr{{\Vector{r}}}

\newcommand\vv{{\Vector{v}}}

\newcommand\vx{{\Vector{x}}}

\newcommand\MATRIX[1]{\mathbf{#1}}

\newcommand\mA{{\MATRIX{A}}}

\newcommand\mH{{\MATRIX{H}}}
\newcommand\mI{{\MATRIX{I}}}

\newcommand\mR{{\MATRIX{R}}}

\newcommand\Prob{\mathbb{P}}
\newcommand\givenp[1][]{{\:#1\vert\:}}

\newcommand\E{\mathbb{E}}

\DeclareMathOperator*{\argmin}{arg\,min}

\usepackage{amsthm}
\newtheorem{theorem}{Theorem}

\newtheorem{lemma}[theorem]{Lemma}
\begin{document}

\title{Antenna Efficiency in Massive MIMO Detection
}

\author{
\IEEEauthorblockN{Ruichen Jiang\IEEEauthorrefmark{1}, Ya-Feng Liu\IEEEauthorrefmark{2}}
\IEEEauthorblockA{\IEEEauthorrefmark{1}Department of Electrical and Computer Engineering, The University of Texas at Austin, TX 78712, USA}
\IEEEauthorblockA{\IEEEauthorrefmark{2}LSEC, ICMSEC, Academy of Mathematics and Systems Science, Chinese Academy of Sciences, Beijing, China}
\small{Email: rjiang@utexas.edu, yafliu@lsec.cc.ac.cn}
}

\maketitle

\begin{abstract}
In this paper, we consider the multi-user detection 
problem in a  
multiple-input multiple-output (MIMO) system, where 
the number of receive antennas at the base station (BS) grows 
infinitely large.
We propose a new performance metric, called antenna efficiency, 
to characterize how fast the 
vector error probability (VEP) decreases as the number of receive antennas increases in the large system limit. 
We analyze the optimal maximum-likelihood (ML) detector 
and the simple zero-forcing (ZF) detector and
prove that their antenna efficiency admits a simple closed form, 
which quantifies 
the impacts of the user-to-antenna ratio, the signal-to-noise ratio (SNR), and the constellation set 
on the VEP. 
Numerical results show that our analysis can well describe the empirical detection error 
performance in a realistic massive MIMO system. 
\end{abstract}


\section{Introduction}
The multiple-input multiple-output (MIMO) detection problem is 
a fundamental problem in modern digital communications and plays a 
pivotal role in various applications \cite{Verdu1998,tse2005}. 
In general, the goal is to recover a vector of transmitted symbols 
from the output of a linear channel corrupted by the 
additive white Gaussian noise. 
In this paper, we will focus on the multi-user MIMO setting, where the 
inputs consist of the symbols transmitted by multiple user terminals (UTs) 
and the outputs consist of the received signals from the multiple antennas at the base station (BS). 

Over the decades, many detection algorithms have been proposed in the literature.
The classic sphere decoding algorithm \cite{Damen2000} 
offers an efficient implementation of  
the maximum-likelihood (ML) detector, which achieves the theoretically optimal 
vector error probability (VEP) under some mild conditions.  
However, its expected complexity grows exponentially as the number of UTs increases \cite{Jalden2005}.
There also exist a large branch of suboptimal algorithms that
trade performance for complexity. 
The incomplete list includes the linear detectors, 
the semidefinite relaxation detectors, and 
the lattice-reduction aided detectors: we refer the readers to 
the survey \cite{fifty_years_of_MIMO} and the references therein. 
Conventionally, the error performance of these detectors is analyzed 
in terms of the diversity order~\cite{tse2005,Winters1994,Jalden2008,Gan2009}, 
which specifies how fast the VEP decays to zero as the signal-to-noise ratio (SNR) tends to infinity. 

Recent years have seen revived interest 
in the signal detection of a large-scale multi-user MIMO system, 
thanks to the great advances in the massive MIMO technology. 
In this emerging scenario, the BS is equipped with hundreds of or even thousands of antennas 
to serve a large number of power-limited UTs simultaneously. 
The traditional error analysis via the diversity order 
is not directly applicable here since it usually employs high-SNR approximations. 
Hence, 
we propose to study the error performance in the large system regime, where 
the SNR per UT is kept fixed but the number of receive antennas at the BS 
grows unbounded \cite{Wagner2012,Hoydis2013}.  
Specifically, we propose a new performance metric termed 
antenna efficiency\footnote{In antenna theory the name ``antenna efficiency" denotes the ratio of the total radiated power to the total input power of an antenna, and we would like to clarify that it is a completely different concept from ours.} 
that measures how much reduction in the VEP can be expected with 
an additional receive antenna asymptotically.  
We analyze the antenna efficiency of 
the optimal ML detector and the simple zero-forcing (ZF) detector and show that  
they admit a simple closed form depending on
the UT-to-antenna ratio, the SNR, and the constellation set.

We adopt the following standard notations in this paper. 
We use $x_i$ to denote the $i$-th entry of a vector $\vx$ and 
$[\mA]_{ij}$ to denote the $(i,j)$-th entry of a matrix $\mA$. 
We use $|\cdot|$ to denote the modulus of a complex number, 
$(\cdot)^\dagger$ to denote the conjugate transpose,
and $\|\cdot\|_2$ to denote the Euclidean norm of a vector.  
The symbol $\mI_m$ denotes the $m\times m$ identity matrix. 
$\E[\cdot]$ denotes the expectation operator, and  
sometimes $\E_{X}[\cdot]$ is used to stress that the expectation is taken 
with respect to the random variable $X$. 
The moment-generating function (MGF) of a random variable 
$X$ is defined as $M_{X}(t)= \E_{X}[\exp(tX)]$. 
We use $\Prob(\cdot)$ and $\Prob(\cdot \givenp \cdot)$ to denote the 
unconditional and conditional probability, respectively. 
We use $\mathcal{CN}(\0, \mR)$ to denote 
the distribution of a circular symmetric complex Gaussian random vector  
with zero mean and covariance matrix $\mR$.

\section{System Model and Problem Formulation}\label{sec:system_model}
\subsection{System Model}
Consider a multi-user MIMO system consisting of $n$ 
UTs with a single antenna
and a BS with $m$ antennas. 
Throughout the paper, we assume that $m\geq n$. 
We consider perfectly synchronized transmissions over flat-fading channels. 
The received signal vector~$\vr \in \mathbb{C}^m$ at the BS is then given by  
\begin{equation}\label{eq:mimo_channel}
    \vr = \mH \vx^*+\vv,
\end{equation}
where 
$\mH\in \mathbb{C}^{m\times n}$ is a complex channel matrix,
$\vx^*\in \mathbb{C}^{n}$ is the vector of transmitted symbols, and 
$\vv\in \mathbb{C}^{m}$ is the noise vector. 
The $j$-th column of $\mH$, denoted by $\vh_j$, represents 
the channel from UT $j$ to the BS. 
We further assume that all entries of $\mH$ 
are independent and identically distributed (i.i.d.) following $\mathcal{CN}(0,1)$, 
and all entries of $\vv$ are i.i.d.~following $\mathcal{CN}(0,\sigma^2)$. 
We define the SNR as the received SNR per UT: 
\begin{equation}\label{eq:SNR}
    \mathrm{SNR} = \frac{\E[\|\mH\vx^*\|_2^2]}{n\E[\|\vv\|_2^2]} = \frac{m\E[\|\vx^*\|_2^2]}{mn \sigma^2}
    = \frac{\E[\|\vx^*\|_2^2]}{n \sigma^2}.
\end{equation} 

The transmitted symbols $x^*_1,\ldots,x^*_n$ are drawn from a constellation set 
$\mathcal{S}$ of size $M$. 
Our results apply to an arbitrary constellation set, including 
the 
$M$-PSK and $M$-QAM constellations. 
In particular,  
the key quantity that plays a crucial role in our analysis is  
the minimum distance of $\mathcal{S}$ defined as 
\begin{equation*}
    d_{\mathrm{min}} = \min_{s,s'\in \mathcal{S},s\neq s'} |s-s'|.
\end{equation*} 
Intuitively, it affects the ``hardness'' of the detection problem: 
a larger $d_{\mathrm{min}}$ implies that the symbols are separated further from each other
and hence easier to distinguish and detect.

\subsection{Two MIMO Detectors}
Given the received signal vector $\vr$ and the channel realization $\mH$, a MIMO detection algorithm 
outputs $\hat{\vx}\in \mathcal{S}^n$ as an estimate of $\vx^*$.
The goal is to minimize the VEP defined as 
$\Prob(\hat{\vx}\neq \vx^*)$.
In this paper, we will focus on two well-studied detectors: the ML detector 
and the ZF detector. 

The ML detector is given by 
\begin{equation}\label{eq:ML}
    \hat{\vx}_{\mathrm{ML}} = \argmin_{\vx\in \mathcal{S}^n} \|\mH\vx-\vr\|_2^2.
\end{equation}
Under the assumption that the UTs choose their transmitted symbols 
from $\mathcal{S}$ uniformly and independently,  
the ML detector is known to be optimal in terms of achieving the minimum possible VEP. 
However, the combinatorial optimization problem in \eqref{eq:ML} is 
strongly NP-hard \cite{Verdu1989} and hence the computational cost of globally solving it can be prohibitively high especially 
when the number of UTs $n$ is large. 
Still, it can serve as a benchmark for other detection algorithms. 

The ZF detector belongs to the family of linear detectors 
and is computationally cheap. 
It first multiplies the received signal vector $\vr$ by the pseudoinverse of $\mH$ to get
\begin{equation}\label{eq:decorrelated}
    \tilde{\vx} = (\mH^\dagger \mH)^{-1}\mH^\dagger \vr,
\end{equation}
and then maps each entry of the decorrelated signal vector $\tilde{\vx}$ to the nearest constellation symbol in $\mathcal{S}$:  
\begin{equation}\label{eq:ZF}
    \hat{x}_{\mathrm{ZF},j} = \argmin_{s\in \mathcal{S}} |s-\tilde{x}_j|,
    \quad j=1,\ldots,n.
\end{equation}
Due to its simplicity, the ZF detector often allows closed-form theoretical results
and is well studied in the literature~\cite{tse2005}. 

\subsection{Antenna Efficiency}
In general, 
the exact VEP of a MIMO detector is intractable  
and it is common to rely on asymptotic analysis.  
A classic performance metric is the diversity order \cite{tse2005} defined by
\begin{equation}\label{eq:diversity_order}
    d = \lim_{\mathrm{SNR}\rightarrow +\infty} -\frac{\log \Prob(\hat{\vx}\neq \vx^*) }{\log \mathrm{SNR}},
\end{equation}
and it characterizes the rate at which the VEP tends to zero in the high SNR regime. 
Specifically, the diversity analysis implies that the VEP of the detector should scale as $\mathrm{SNR}^{-d}$ at high SNR.
It is known that the ML detector enjoys diversity order of $m$ \cite{Nee2000,Zhu2002}, 
while the ZF detector only achieves diversity order of $m-n+1$ \cite{Winters1994}. 

In this paper, we take an alternative view and study 
how the VEP behaves in the large system limit, where the number of antennas $m$ grows  
infinitely large while the ratio $n/m$ tends to $\delta\in [0,1]$. 
This includes the special case where the number of UTs $n$ remains fixed by letting $\delta=0$.  
Analogous to the definition of diversity order in \eqref{eq:diversity_order}, we define
\begin{equation}\label{def:ae}
    f = \lim_{m\rightarrow +\infty,\; n/m\rightarrow \delta}
    -\frac{\log \Prob(\hat{\vx}\neq \vx^*)}{m}.
\end{equation}
Roughly speaking, with \eqref{def:ae} we shall have $\Prob(\hat{\vx}\neq \vx^*) \approx e^{-fm}$, 
which means that each additional antenna at the BS will bring $4.34f$ dB decrease in the VEP. 
In this sense, $f$ characterizes how efficiently we can reduce the 
VEP by increasing the number of antennas, and hence we name it as 
\emph{antenna efficiency}. 
As it will become clearer, the antenna efficiency in \eqref{def:ae} is a function of the ratio $\delta$, the noise variance $\sigma^2$, and the constellation set $\mathcal{S}$.  
In the following, 
we will quantify such dependence and give the antenna efficiency of 
the ML and ZF detectors in simple closed form.
 
It is worth mentioning that some researchers have also considered 
MIMO detection in the large system limit \cite{Tanaka2002,Jeon2015,Thrampoulidis2018}, 
but with different channel models from ours. More specifically, 
they assumed that the entries of $\mH$ are i.i.d.~real/complex zero-mean Gaussian random variables with  
variance either $1/n$ 
(i.e., the total transmitted power is fixed)
or $1/m$ (i.e., the received power per UT is fixed). 
This difference leads to a very different asymptotic behavior:
instead of tending to zero at an exponential rate, 
the VEP is shown to converge to a nonzero limit when $m$ goes to infinity. 
We justify our system model in twofold. 
First, for a multi-user MIMO system, 
it is reasonable to assume that the UTs have individual power supplies 
and the captured energy at the BS increases linearly with the number of 
antennas $m$.   
Such assumptions are also widely adopted in the literature \cite{Wagner2012,Hoydis2013}.  
Second, 
our analysis appears more elementary, whereas the existing works involve more sophisticated tools 
such as the replica method \cite{Tanaka2002} and the Gaussian comparison inequalities \cite{Thrampoulidis2018}. 
This simplicity also enables us to generalize our results to the per-user spatially correlated channel model as in \cite{Wagner2012} 
and possibly other problems in the massive MIMO system such as symbol-level precoding~\cite{Alodeh2018}, which  
we will put as future works. 

\section{Antenna Efficiency of ML Detector}\label{sec:ML}
In this section, we characterize and prove the antenna efficiency of 
the ML detector.
\begin{theorem}\label{thm:ML}
Consider the MIMO system in \eqref{eq:mimo_channel}. Assume that the entries of 
$\mH$ are i.i.d.~following $\mathcal{CN}(0,1)$, the entries of $\vv$ are i.i.d.~following $\mathcal{CN}(0,\sigma^2)$, 
and the entries of $\vx^*$ are drawn uniformly and independently from the constellation set $\mathcal{S}$ with  
minimum distance $d_{\mathrm{min}}$.   
Then for the ML detector in~\eqref{eq:ML}, its antenna efficiency is given by   
    \begin{equation}\label{eq:ae_ML}
        f_{\mathrm{ML}} = \log  \left(  1+\frac{d_{\mathrm{min}}^2}{4{\sigma}^2}\right).
    \end{equation}
\end{theorem} 

Theorem~\ref{thm:ML} offers a simple formula for $f_{\mathrm{ML}}$ and 
shows that it 
is determined by the quantity ${d^2_{\mathrm{min}}}/{(4\sigma^2)}$, which 
can be regarded as the effective detection SNR. 
Also, note that $f_{\mathrm{ML}}$
is independent of the UT-to-antenna ratio~$\delta$. 
Since a larger $\delta$ means more UTs in the system and hence 
higher interference,  
we deduce that the ML detector is able to suppress    
the multi-user interference effectively.   

To prove Theorem~\ref{thm:ML}, we first derive a lower bound on the VEP by assuming no multi-user interference,  
and then provide a matching upper bound by using the union bound. 
\subsection{No-Interference Lower Bound}\label{subsec:no_inter}
Note that $\Prob(\hat{\vx} \neq \vx^*) \geq \Prob(\hat{x}_j \neq x^*_j)$ for all 
$j=1,\ldots,n$. Hence, in the following we will lower bound 
$\Prob(\hat{x}_j \neq x^*_j)$, which in turn 
results in a lower bound on the VEP.

Without loss of generality, let $j=1$. 
To derive a lower bound, imagine that all
the transmitted symbols $x^*_2,\ldots,x^*_n$ except $x^*_1$ are known. 
Equivalently,  
this can be interpreted as the idealistic assumption that 
the interference caused by transmissions from other UTs is perfectly cancelled out. 
In this case, 
the detection of $x^*_1$ reduces to a $1\times m$ single-input multiple-output detection problem
\begin{equation}\label{eq:simo}
    \tilde{\vr}=\vh_1 x^*_1+\vv,
\end{equation}
where $\tilde{\vr} = \vr - \sum_{j=2}^n \vh_j x^*_j$. Furthermore, 
it can be shown that the vector detection in \eqref{eq:simo} is equivalent to 
a scalar detection problem~\cite[Sec. A.2.3]{tse2005}:
\begin{equation*}
    \frac{\vh_1^\dagger \tilde{\vr}}{\|\vh_1\|_2^2} = x_1^*+ \frac{\vh_1^\dagger \vv}{\|\vh_1\|_2^2},
\end{equation*}
where the variance of the equivalent noise is given by $\sigma^2/\|\vh_1\|_2^2$. 
It follows from standard results on scalar detection~\cite[Sec. 4.2]{Proakis2008} that
\begin{equation}\label{eq:condition_MFB}
    \Prob(\hat{x}_1 \neq x_1^*\givenp \vh_1) \geq \frac{2}{M}Q\left(\frac{d_{\mathrm{min}}\|\vh_1\|_2}{\sqrt{2}\sigma}\right),
\end{equation}
where the $Q$-function is $ Q(x) = \frac{1}{\sqrt{2\pi}}\int_{x}^{\infty} \exp\left(-\frac{u^2}{2}\right)du$.

It remains to take the expectation over $\vh_1$. Note that $2\|\vh_1\|^2_2$ is 
a chi-square random variable with $2m$ degrees of freedom, i.e., $2\|\vh_1\|_2^2\sim \chi^2_{2m}$. 
Hence, its MGF can be computed as $M_{\|\vh_1\|_2^2}(t)=(1-t)^{-m}$. 
Using the Craig's representation of $Q(x)$ \cite{Craig1991}, i.e., 
\begin{equation*}
    Q(x) = \frac{1}{\pi}\int_{0}^{{\pi}/{2}} \exp \left( -\frac{x^2}{2\sin^2(\theta)} \right) d\theta,
\end{equation*}
and exchanging the order of expectation and integration, 
from~\eqref{eq:condition_MFB} we get 
\begin{align*}
    \Prob(\hat{x}_1 \neq x_1^*) &\geq 
    \frac{2}{\pi M} \int_{0}^{{\pi}/{2}} \E_{\vh_1} \left[\exp \left( -\frac{d_{\mathrm{min}}^2\|\vh_1\|_2^2}{4\sigma^2\sin^2(\theta)} \right)\right]  d\theta \nonumber\\
    &= \frac{2}{\pi M} \int_{0}^{{\pi}/{2}} M_{\|\vh_1\|_2^2}\left(-\frac{d_{\mathrm{min}}^2}{4\sigma^2\sin^2(\theta)}\right) d\theta \nonumber\\
    &= \frac{2}{\pi M} \int_{0}^{{\pi}/{2}} \left(1+\frac{d_{\mathrm{min}}^2}{4\sigma^2\sin^2(\theta)}\right)^{-m} d\theta \nonumber\\
    &\geq \frac{2}{\pi M}\left(  1+\frac{d_{\mathrm{min}}^2}{4{\sigma}^2}\right)^{-m} \int_{0}^{{\pi}/{2}} \sin^{2m}(\theta) d\theta.   
\end{align*}
Combining this with Wallis' formula \cite{Kazarinoff1956}    
\begin{equation*}
    \int_{0}^{\pi/2} \sin^{2k}(\theta) d\theta \geq \frac{\sqrt{\pi}}{2 \sqrt{k+\frac{1}{2}}},\quad \forall\;k \geq 1,
\end{equation*}
yields
\begin{equation*}
    \Prob(\hat{x}_1 \neq x_1^*) \geq \frac{1}{\sqrt{\pi(m+\frac{1}{2})}M}\left(  1+\frac{d_{\mathrm{min}}^2}{4{\sigma}^2}\right)^{-m}.
\end{equation*}
It follows immediately from the definition in \eqref{def:ae} that 
\begin{equation*}
    f_{\mathrm{ML}} \leq 
    \log  \left(  1+\frac{d_{\mathrm{min}}^2}{4{\sigma}^2}\right).
\end{equation*} 
\subsection{Union Upper Bound}
From \eqref{eq:ML}, we can see that the ML detector fails to recover the 
true transmitted symbol vector only if there exists $\vx'\in \mathcal{S}^n$ different from $\vx^*$ such that 
$\|\mH\vx^*-\vr\|_2\geq \|\mH\vx'-\vr\|_2$. 
Therefore, we define the pairwise error probability as 
\[\Prob(\vx^*\rightarrow \vx'):=\Prob(\|\mH\vx^*-\vr\|_2\geq \|\mH\vx'-\vr\|_2),\]
and the union bound leads to   
\begin{equation}\label{eq:union_bound}
    \Prob(\hat{\vx}_{\mathrm{ML}}\neq \vx^*) 
    \leq \sum_{\vx'\in \mathcal{S}^n,\vx'\neq \vx^*}\Prob(\vx^*\rightarrow \vx').
\end{equation}
Now we take a closer look at the pairwise error probability. Conditioned on 
$\mH$, it corresponds to a detection problem with a binary symbol set $\{\mH\vx^*,\mH\vx'\}$. Hence, 
we have 
\begin{align}
    \Prob(\vx^*\rightarrow \vx' \givenp \mH) &=   Q\left( \frac{\|\mH(\vx^*-\vx')\|_2}{\sqrt{2}\sigma} \right)  \nonumber \\
                                &\leq \frac{1}{2} \exp\left( -\frac{1}{4\sigma^2}\|\mH(\vx^*-\vx')\|_2^2 \right), \label{eq:bound_pep} 
\end{align}
where we used the standard inequality 
\begin{equation}\label{eq:Q_upper_bound}
    Q(x) \leq \frac{1}{2}e^{-\frac{x^2}{2}}.
\end{equation} 

To average over the randomness of $\mH$, we write  
\begin{equation*}
    \mH(\vx^*-\vx') = \sum_{j=1}^n (x^*_j-x'_j)\vh_j,
\end{equation*}
where by our assumption $\vh_1,\ldots,\vh_n$ are i.i.d.~following $\mathcal{CN}(\0,\mI_m)$. 
Hence, $\mH(\vx^*-\vx')$ is a complex Gaussian random vector with zero mean and covariance matrix being 
$\|\vx^*-\vx'\|_2^2 \mI_m$, 
which implies that $\frac{2\|\mH(\vx^*-\vx')\|_2^2}{\|\vx^*-\vx'\|_2^2}\sim \chi_{2m}^2$.   
By this and using its MGF, we get from \eqref{eq:bound_pep} that 
\begin{equation}\label{eq:PEP}
    \Prob(\vx^*\rightarrow \vx') \leq \frac{1}{2} \left(1+\frac{\|\vx^*-\vx'\|_2^2}{4\sigma^2}\right)^{-m}.
\end{equation}

Now we distinguish two cases: $n$ remains bounded as $m$ tends to infinity, 
or $n$ also grows unbounded. 
In the first case, 
since $\vx'$ is different from $\vx^*$ in at least one entry, 
we have $\|\vx^*-\vx'\|_2 \geq d_{\mathrm{min}}$. 
Then \eqref{eq:union_bound} and \eqref{eq:PEP} lead to 
\begin{equation*}
    \Prob(\hat{\vx}_{\mathrm{ML}}\neq \vx^*) \leq \frac{M^n-1}{2}\left(1+\frac{d_{\mathrm{min}}^2}{4\sigma^2}\right)^{-m}.
\end{equation*}
Since both $M$ and $n$ are upper bounded by some absolute constant independent of $m$, 
we arrive at 
\begin{equation}\label{eq:f_ML_lower_bound}
    f_{\mathrm{ML}} \geq \log  \left(  1+\frac{d_{\mathrm{min}}^2}{4{\sigma}^2}\right).
\end{equation}

The second case requires more refined analysis. If $\vx'$ is different from 
$\vx^*$ in $k$ entries, then $\|\vx^*-\vx'\|^2_2\geq kd_{\mathrm{min}}^2$. 
By grouping $\vx'$ according to the number of incorrect entries, 
the union bound \eqref{eq:union_bound} and \eqref{eq:PEP} result in 
\begin{equation}\label{eq:union_bound_binomial}
    \Prob(\hat{\vx}_{\mathrm{ML}}\neq \vx^*) \leq \frac{1}{2}\sum_{k=1}^n 
    \binom{n}{k} (M-1)^{k}\left(1+\frac{k d^2_{\mathrm{min}}}{4\sigma^2}\right)^{-m}.
\end{equation}
Recall that we always assume $m\geq n$. 
Hence, when $n$ is large enough, we can expect that the first summand (i.e., the term corresponding to $k=1$) in the right-hand side of 
\eqref{eq:union_bound_binomial}  will dominate. 
The following lemma formalizes this observation.  
\begin{lemma}\label{lem:union_upper_bound}
    Let $\rho=d_{\mathrm{min}}^2/(4\sigma^2)$. 
    If $n$ is larger than
    \begin{equation*}
        \begin{aligned}
            \max\biggl\{\max\Bigl\{4(M-1),2\sqrt{2e(M-1)}&\Bigr\}\biggl(1+\frac{1}{\rho}\biggr), \\
            \frac{1}{2}\biggl(2+\frac{1}{\rho}&\biggr)^2, \frac{2\sqrt{2}+2}{\rho}\biggr\},
        \end{aligned}
    \end{equation*}  
    then we have 
    \begin{equation*}
        \Prob(\hat{\vx}_{\mathrm{ML}}\neq \vx^*) 
        \leq \frac{1}{2}
        \left(M+\frac{(M-1)^2}{2\log^2\left(\frac{1+2\rho}{1+\rho}\right)}\right)
    n(1+\rho)^{-m}.
    \end{equation*}
\end{lemma}
Similarly, by taking the limit as in \eqref{def:ae}, we also obtain the lower bound in \eqref{eq:f_ML_lower_bound} 
from Lemma~\ref{lem:union_upper_bound}. 
\section{Antenna Efficiency of ZF Detector}
In this section, we will characterize and prove the 
antenna efficiency of the ZF detector. 
\begin{theorem}\label{thm:ZF}
Consider the MIMO system in \eqref{eq:mimo_channel} with the same assumptions as in Theorem~\ref{thm:ML}. 
Then for the ZF detector defined by \eqref{eq:decorrelated} and \eqref{eq:ZF}, its antenna efficiency 
is given by
    \begin{equation}\label{eq:ae_ZF}
        f_{\mathrm{ZF}} = (1-\delta) \log  \left(  1+\frac{d_{\mathrm{min}}^2}{4{\sigma}^2}\right).
    \end{equation}
\end{theorem}

Compared with \eqref{eq:ae_ML},  
the suboptimality of the ZF detector is reflected in the coefficient $1-\delta$. 
On the one hand, when the number of antennas $m$ is much larger than the number of UTs $n$, 
the ratio $\delta$ is close to 0 and the ZF detector should achieve near-optimal antenna efficiency 
as the ML detector. 
On the other hand, when $n$ scales linearly with $m$, the performance of 
the ZF detector deteriorates due to the multi-user interference. 

Similar to Section~\ref{sec:ML}, we prove Theorem~\ref{thm:ZF} by deriving matching lower and upper bounds 
on the VEP. We first write the decorrelated signal vector $\tilde{\vx}$ in \eqref{eq:decorrelated} as
\begin{equation}\label{eq:decorrelated_channel}
    \tilde{\vx} = (\mH^\dagger \mH)^{-1}\mH^\dagger \vr = \vx^*+ \tilde{\vv},
\end{equation}
where $\tilde{\vv} = (\mH^\dagger \mH)^{-1}\mH^\dagger\vv$. 
We can see from \eqref{eq:ZF} and \eqref{eq:decorrelated_channel} that the ZF detector transforms the MIMO channel into $n$ parallel scalar channels and 
decides the transmitted symbol of each UT separately. 
Hence, it is easier to analyze the symbol error probability $\Prob(\hat{x}_{\mathrm{ZF},j}\neq x^*_j)$,
which is related to the VEP by 
\begin{equation}\label{eq:SEP_VEP}
    \Prob(\hat{x}_{\mathrm{ZF},j}\neq x^*_j) \leq \Prob(\hat{\vx}_{\mathrm{ZF}}\neq \vx^*) \leq \sum_{j=1}^n \Prob(\hat{x}_{\mathrm{ZF},j}\neq x^*_j).
\end{equation}
Moreover, since the UTs' channels are statistically equivalent, we have 
$\Prob(\hat{x}_{\mathrm{ZF},j}\neq x^*_j)=\Prob(\hat{x}_{\mathrm{ZF},1}\neq x^*_1)$ 
for all $j=1,\ldots,n$ and \eqref{eq:SEP_VEP} further reduces to 
\begin{equation}\label{eq:SEP_VEP_simple}
    \Prob(\hat{x}_{\mathrm{ZF},1}\neq x^*_1) \leq \Prob(\hat{\vx}_{\mathrm{ZF}}\neq \vx^*) \leq 
    n\Prob(\hat{x}_{\mathrm{ZF},1}\neq x^*_1).
\end{equation}

Conditioned on $\mH$, the equivalent noise vector $\tilde{\vv}$ follows a 
complex Gaussian distribution with zero mean and covariance matrix being
\begin{equation*}
    \E\left[\tilde{\vv}\tilde{\vv}^\dagger\right] =\sigma^2 (\mH^\dagger \mH)^{-1}\mH^\dagger \mH (\mH^\dagger \mH)^{-1} = \sigma^2 (\mH^\dagger \mH)^{-1}.
\end{equation*}
Hence, the first UT sees a scalar channel   
with the equivalent noise variance $\sigma^2 [(\mH^\dagger \mH)^{-1}]_{11}$.  
Following standard results on scalar detection \cite{Proakis2008}, we have 
\begin{align}
    \Prob(\hat{x}_{\mathrm{ZF},1} \neq x_1^*\givenp \mH) &\geq 
    \frac{2}{M}Q\left(\frac{d_{\mathrm{min}}\sqrt{\gamma_1}}{\sqrt{2}\sigma}\right), \label{eq:ZF_lb}\\
    \Prob(\hat{x}_{\mathrm{ZF},1} \neq x_1^*\givenp \mH) &\leq 
    (M-1)Q\left(\frac{d_{\mathrm{min}}\sqrt{\gamma_1}}{\sqrt{2}\sigma}\right),\label{eq:ZF_ub}
\end{align}
where $\gamma_1 = 1/[(\mH^\dagger \mH)^{-1}]_{11}$. 

Moreover, it can be shown that  $2\gamma_1 \sim \chi_{2(m-n+1)}^2$~\cite{Jiang2011}. 
Using \eqref{eq:ZF_lb} and following similar arguments as in Section~\ref{subsec:no_inter}, we get 
\begin{equation*}
    \Prob(\hat{x}_{\mathrm{ZF},1} \neq x_1^*) \geq \frac{1}{\sqrt{\pi\left(m-n+\frac{3}{2}\right)}M}\left(  1+\frac{d_{\mathrm{min}}^2}{4{\sigma}^2}\right)^{-m+n-1}\!.
\end{equation*}
Together with \eqref{eq:SEP_VEP_simple}, it implies that 
\begin{equation*}
    f_{\mathrm{ZF}} \leq (1-\delta)  \log  \left(  1+\frac{d_{\mathrm{min}}^2}{4{\sigma}^2}\right).
\end{equation*} 
On the other hand, combining \eqref{eq:Q_upper_bound} and \eqref{eq:ZF_ub} yields  
\begin{equation*}
    \Prob(\hat{x}_{\mathrm{ZF},1} \neq x_1^*) \leq \frac{M-1}{2} \left(  1+\frac{d_{\mathrm{min}}^2}{4{\sigma}^2}\right)^{-m+n-1}.
\end{equation*}
Together with \eqref{eq:SEP_VEP_simple}, this leads to  
\begin{equation*}
    f_{\mathrm{ZF}} \geq (1-\delta)  \log  \left(  1+\frac{d_{\mathrm{min}}^2}{4{\sigma}^2}\right).
\end{equation*}

\section{Numerical Results}
\begin{figure}[!t]
    \centering
    \includegraphics[width=0.96\linewidth, clip=true]{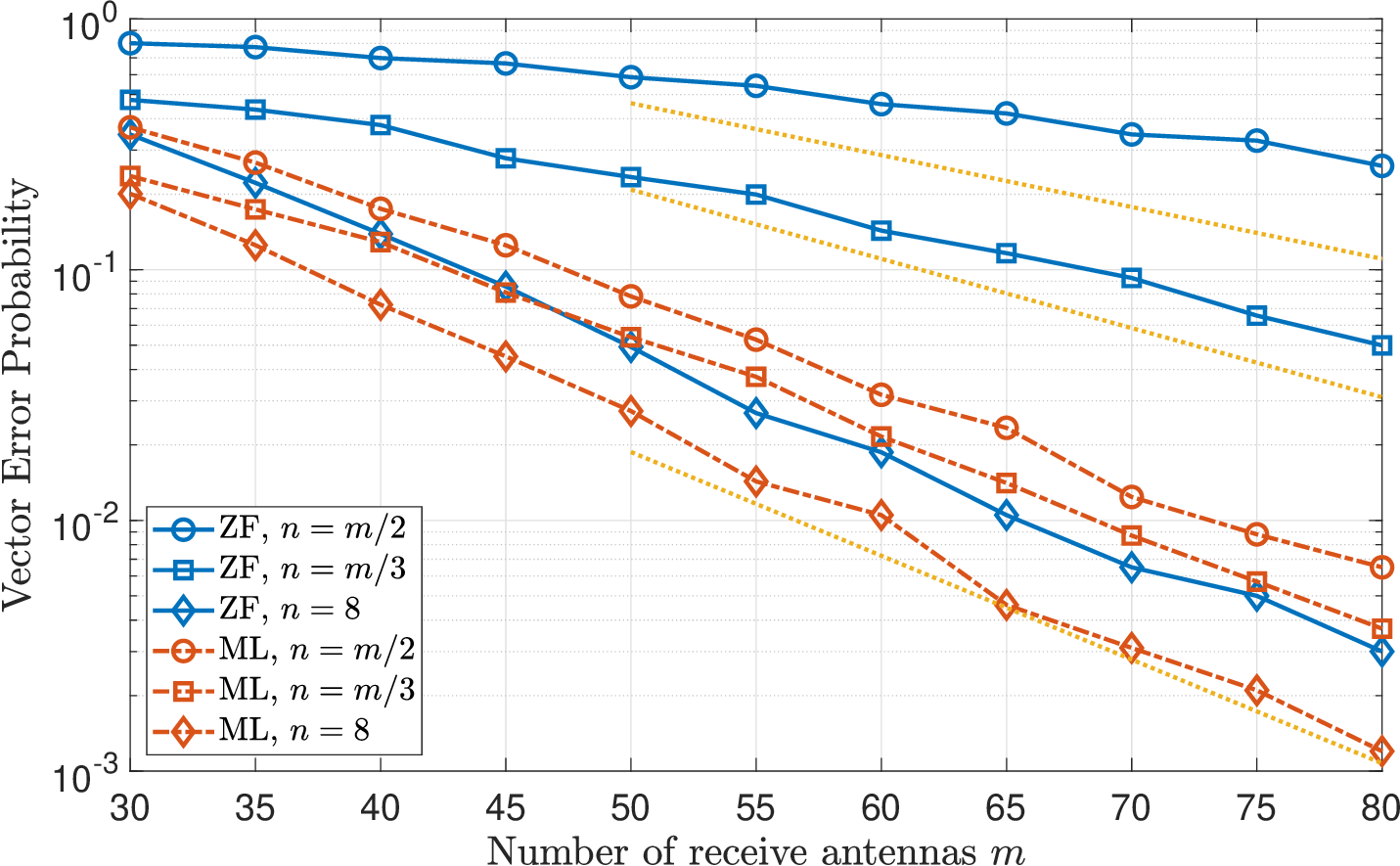}
    \caption{The VEP versus the number of antennas $m$ where $\mathrm{SNR}=0$ dB and 
    16-QAM is used.}\label{fig:varying_delta}
\end{figure}

\begin{figure}[!t]
    \centering
    \includegraphics[width=0.96\linewidth, clip=true]{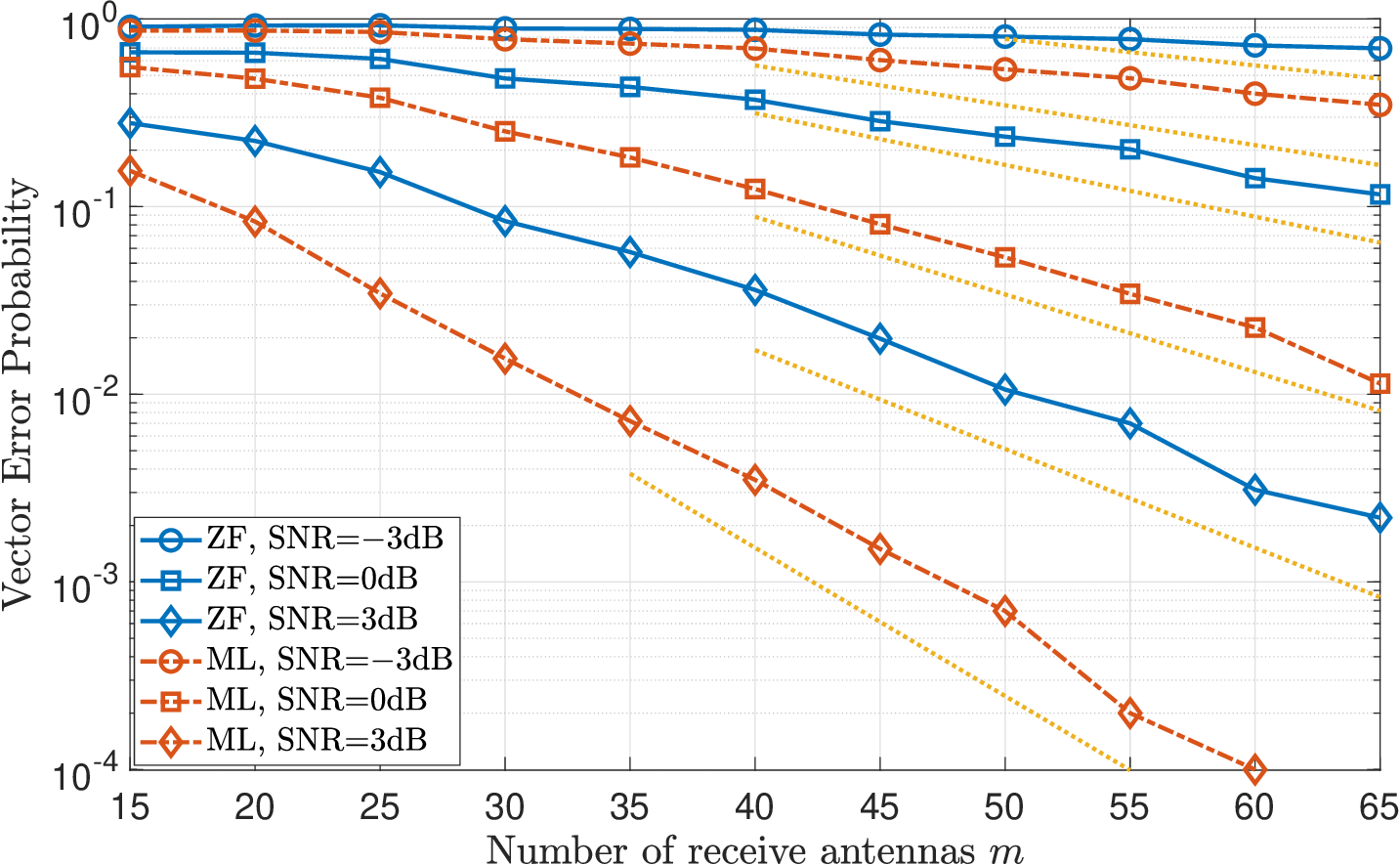}
    \caption{The VEP versus the number of antennas $m$ where $\delta=1/3$ and 
    16-QAM is used.}\label{fig:varying_SNR}
\end{figure}

\begin{figure}[!t]
    \centering
    \includegraphics[width=0.96\linewidth, clip=true]{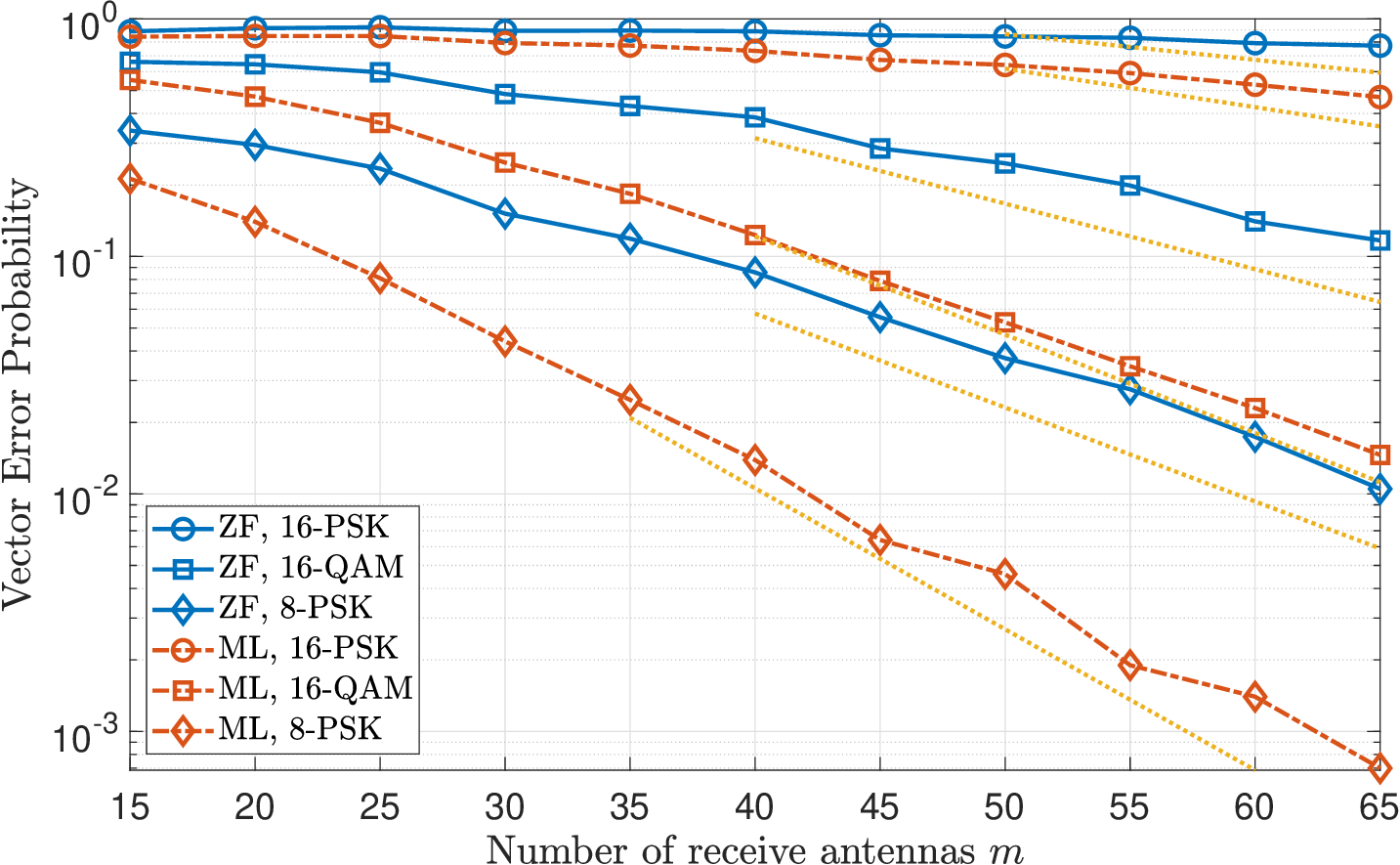}
    \caption{The VEP versus the number of antennas $m$ where $\mathrm{SNR}=0$ dB and $\delta=1/3$.}\label{fig:varying_modulation}
\end{figure}

In this section, we present some numerical results to validate our analysis. 
In our simulations, the constellation set $\mathcal{S}$ is normalized such that 
$\E[|x^*_j|^2]=1$ for all $j$ and hence  
$\mathrm{SNR} = 1/\sigma^2$ (cf.~\eqref{eq:SNR}). 
We test the impacts of the UT-to-antenna ratio $\delta$, the SNR, and the constellation set $\mathcal{S}$ on 
the VEP. 
The results are shown in Figs.~\ref{fig:varying_delta}--\ref{fig:varying_modulation} respectively,
where each data point is the average of 10,000 random instances.

By the definition in \eqref{def:ae}, we can view the antenna efficiency 
as the negative slope of the VEP (in the log scale) 
versus the number of receive antennas $m$ in the large system limit. 
Fig.~\ref{fig:varying_delta} shows that while 
a smaller ratio $\delta$ leads to a steeper slope for the ZF detector, 
it does not affect the ML detector in terms of the asymptotic performance. 
Also, when the number of UTs is fixed, which corresponds to $\delta=0$,
the ZF detector can achieve 
the optimal antenna efficiency as predicted by our analysis. 
On the other hand, from Fig.~\ref{fig:varying_SNR} and Fig.~\ref{fig:varying_modulation}, 
we can see that both the ZF and ML detectors can benefit from a higher SNR or 
a constellation set with a larger minimum distance. 

To compare our theoretical analysis 
with the empirical results, in Figs.~\ref{fig:varying_delta}--\ref{fig:varying_modulation} 
we also plot the dotted lines with the slopes given by \eqref{eq:ae_ML} or \eqref{eq:ae_ZF}
alongside the corresponding VEP curves. 
From the figures, we can observe that the empirical VEP agrees well with 
our analysis across various settings.
Despite the asymptotic nature of our analysis,  
the VEP exhibits an exponential decay with a moderate number of antennas.


\bibliographystyle{IEEEtran}

\end{document}